\documentclass[12pt]{elsarticle}
\usepackage{amsmath}
\usepackage{amssymb}
\usepackage{graphicx}
\usepackage{placeins}
\usepackage{pdfpages,todonotes}

\usepackage{natbib}
\setcitestyle{authoryear,open={(},close={)}}

\makeatletter
\usepackage{comment}
\usepackage{float}
\usepackage{booktabs} \newcommand{\ra}[1]{\renewcommand{\arraystretch}{#1}}
\makeatother

\makeatletter
\def\ps@pprintTitle{%
   \let\@oddhead\@empty
   \let\@evenhead\@empty
   \def\@oddfoot{\reset@font\hfil\thepage\hfil}
   \let\@evenfoot\@oddfoot
}
\makeatother

\begin{document}

\begin{frontmatter}

\title{Estimating TVP-VAR models with time invariant long-run multipliers}
\author[label1,label2]{Denis Belomestny}
\author[label1,label2]{Ekaterina Krymova\corref{cor1}}
\author[label2,label3]{Andrey Polbin}
\address[label1]{University  of Duisburg-Essen, Essen, Germany}
\address[label2]{Russian  Academy of National Economy and Public Administration, Moscow, Russia }
\address[label3]{Gaidar Institute for Economic Policy, Moscow, Russia}

\cortext[cor1]{
Corresponding author. E-mail address: ekaterina.krymova@epfl.ch\\
Present address: SDSC EPFL, Lausanne, Switzerland}

\newpage

\begin{abstract}
The main goal of this paper is to develop a methodology for estimating time varying parameter vector auto-regression (TVP-VAR) models
with a time-invariant long-run relationship between endogenous variables and changes in exogenous variables. We propose a Gibbs sampling scheme for estimation of model parameters as well as time-invariant long-run multiplier parameters. Further we demonstrate the applicability of the proposed method by analyzing examples of the Norwegian and Russian economies based on the data on  real GDP, real exchange rate and real oil prices.
Our results show that incorporating the time invariance constraint on the long-run multipliers in TVP-VAR model helps to significantly improve the  forecasting performance.
\end{abstract}


\end{frontmatter}

\section{Introduction}
During the past two decades time-varying parameter estimation became very popular in macroeconomic modeling. The existing literature provides strong evidence for a time  varying  behavior of  volatility \citep{primiceri2005time,justiniano2008time,mcconnell2000output},  long-run economic growth \citep{kim1999has,cogley2005fast,antolin2017tracking},   trend inflation   \citep{cogley2008trend,stock2007has,clark2014evaluating},    inflation persistence \citep{cogley2010inflation,kang2009changes}, oil price persistence  \citep{kruse2019time},  dependence of main macroeconomic variables on oil prices  \citep{baumeister2013time,chen2009oil,cross2017relationship,riggi2015time}.

After seminal papers \citep{primiceri2005time,del2015time,cogley2005drifts} Bayesian time-varying parameter vector autoregression (TVP-VAR) model  with stochastic volatility became one of the main modeling tools to capture  temporary changes in relations between the variables. Time-varying parameters are believed to follow simple stochastic processes, parameters of which are estimated with the help of Monte Carlo techniques  \citep[see][]{gelfand1990sampling,gelfand1991gibbs,carter1994gibbs}. As demonstrated in   \citep{koop2013large,d2013macroeconomic,clark2015macroeconomic} Bayesian TVP-VAR models could be used for forecasting. Nevertheless,  Bayesian TVP-VAR models have not yet become an ubiquitous forecasting tool due to a large number of  parameters to estimate.
\par
In this paper we consider models with changing in time parameters motivated by a change in economic policy regimes.  According to Lucas critique \citep{lucas1976econometric} rational economic agents take the  structural changes in economy  into account when making decisions. Therefore changes in economic policy should lead to the changes in parameters of such non-structural  models as, for example, large macroeconometric models consisting of simultaneous equations or vector autoregression models.  In a series of papers  the high volatility of US macroeconomic indicators  is related to poor monetary policy performance at the time before Paul Volcker became chairman of the Fed \citep{clarida2000monetary,judd1998taylor,lubik2004testing,mavroeidis2010monetary}.  However, empirical evidence for this hypothesis on the basis of time varying parameter models is controversial. \cite{primiceri2005time} proposed TVP-VAR model and developed a Bayesian method to estimate model parameters.  An example of TVP-VAR modelling of US economy failed to demonstrate the changes in the monetary policy transmission. Along with that \citep{cogley2001evolving,cogley2005drifts,canova2015estimating,gambetti2008structural} provided empirical  evidences of a notable change in the monetary policy transmission mechanism  using TVP-VAR and in \citep{sims2006were} with the help of Markov switching VAR model.
At the same time there is a strong empirical evidence in favour of a nominal exchange rate regime influence on the business cycle performance of  developing countries. A floating exchange rate has a
stabilizing effect on the output under the influence of terms-of-trade shocks. The latter was shown by \cite{broda2004terms} with the help of VAR methods and by  \citep{edwards2005flexible} using panel regression techniques.
In addition, exchange rate regimes in developing countries demonstrate changeable behavior \citep{levy2005classifying}. Thus TVP-VAR models are promising for modeling of economies under exchange rate regime shifts.
\par
We aim to analyze econometric models with time-varying short-term and invariant long-term relationships in order to describe economic system whose cross-correlation relationships change due to changes in the monetary policy and exchange rate regimes. The long-term assumptions arise from a classical hypothesis of the long-run money neutrality. Empirical support in favour of this hypothesis is exhaustively documented in the literature, we cite here only \citep{fisher1993long,king1992testing,weber1994testing}, see also references therein.
Furthermore, the long-run neutrality of monetary policy shocks is a typical assumption in estimation of SVAR models \citep{altig2011firm,canova2015estimating,peersman2005caused}.
We also propose a methodology for estimation of TVP-VAR models with time-invariant long-run relations of endogenous variables to changes in exogenous variables. VAR model with exogenous variables (VARX) is one of the main methods for describing the dynamics of small open economies \citep{cushman1997identifying,fernandez2017world,uribe2006country}.
Natural candidates for exogenous variables in VARX models are oil prices, terms of trade, world interest rates, external demand and many others. 
Hence, the proposed methodology may find application in numerous practical examples.
\par
The paper is organized as follows. Section 2 describes a new methodology for modeling of economy  
based on TVP-VAR model
\citep{primiceri2005time} incorporating non-zero long-run restriction (time invariance of  long-run multipliers). We estimate long-run multipliers within a  Monte Carlo procedure. Section 3 describes a particular case of  modeling  with  real GDP, real exchange rate as endogenous and real oil prices as exogenous variable for an oil exporting country. Section 4 contains the results of estimation of the model on the Norwegian and Russian datasets. We demonstrate the model's forecasting performance in comparison with classical VARX and a modification of TVP-VAR with exogenous variables. Our results show that the time invariance constraint for long-run multiplier brings significantly improvement of TVP-VAR model performance in terms of forecasting accuracy.

 \section{Constrained TVP-VAR}

We recall first a framework of  time varying parameter vector auto-regression model (TVP-VAR)    of \cite{primiceri2005time,del2015time}, where the endogenous time series vector $y_t\in\mathbb{R}^n$ is modeled by the  following measurement equation 
\begin{equation}
y_{t}=c_{t}+B_{1,t}y_{t-1}+\ldots+B_{k,t}y_{t-k}+u_{t},\quad t=1,\ldots,T,\label{eq:primiceri}
\end{equation}
where $B_{i,t},$ $i=1,\ldots,k,$ are $n\times n$ matrices of time varying
coefficients, a random vector $u_{t}\in\mathbb{R}^{n}$ contains
heteroskedastic unobserved shocks with a covariance matrix $\Omega_{t}$. The covariance matrix $\Omega_{t}$ is defined via a decomposition
$$
A_{t}\Omega_t A_{t}^{\top}=\Sigma_{t}\Sigma_{t}^{\top},
$$
where $A_{t}$ is a lower triangle matrix and
$\Sigma_{t}=\mathrm{diag}(\sigma_{1,t},\ldots,\sigma_{n,t})$
is a diagonal matrix.   Then it follows that
\begin{equation}\label{vol_process}
u_t = A^{-1}_t \Sigma_t e_t,
\end{equation}
where $e_t \in\mathbb{R}^{n}$ is a vector with independent standard Gaussian components.
In \cite{primiceri2005time,del2015time} a Bayesian approach was used for statistical inference in this model.
\par
Here we present an extension of the model  \eqref{eq:primiceri}.  In particular, we introduce an exogenous variable
$x_{t}$ and a long run constraint on the VAR coefficients.  Our generalized  TVP-VAR model for the exogenous variables  reads as
\begin{equation}
y_{t}=c_{t}+B_{1,t}y_{t-1}+\ldots+B_{k,t}y_{t-k}+\sum_{i=0}^{k}D_{i,t}x_{t-i}+u_{t},\quad t=1,\ldots,T,\label{eq:exo_primiceri}
\end{equation}
where  $D_{i,t}\in\mathbb{R}^{n}$ are $n$-dimensional time varying 
vectors of coefficients  and $c_{t}$ is  a  $n$-dimensional time-varying intercept term. Note that the exogenous time series enter the right hand side of  \eqref{eq:exo_primiceri} with zero lag. We restrict ourselves for simplicity to the case of one exogenous variable. 
Our goal is to develop a Bayesian estimation procedure for the  extended model with exogenous variables \eqref{eq:exo_primiceri} under
the following long-run time invariant constraints on the vectors of coefficients
$B_{i,t}$ and $D_{i,t}$
\begin{equation}
\theta=\left[I_{n}-\sum_{j=1}^{k}B_{j,t}\right]^{-1}\sum_{i=0}^{k}D_{i,t},\label{eq:constraint}
\end{equation}
where  $\theta\in\mathbb{R}^{n}$ is a constant multiplier parameter.
Thus we impose  condition that the shocks in the exogenous variable lead to the same long-run response in the endogenous vector independently of the time when the shock occurs. As was discussed in introduction, such modelling approach  can be appropriate for economic systems whose cross-correlation relationships change due to changes in the monetary policy and exchange rate regimes if the hypothesis of the long-run neutrality of money holds. In the modern New Keynesian models the particular form of the monetary policy rule matters for the shape of transition path from one long-run equilibrium to another, however, the influence of the monetary policy rule on the long-run equilibrium is usually absent.

\subsection{Bayesian inference\label{sec:Bayesian-inference-with}}

In \cite{primiceri2005time,del2015time} the coefficients of the model (\ref{eq:primiceri}) were
modeled in the following way. Let all the vectors $B_{i,t}$, $i=1,\dots,k$ be stacked into a vector $B_t$ \textcolor{black}{of length} $k\cdot n$, let $\alpha_{t}$ be a vector of non-zero
and non-one elements of the matrix $A_{t}$ (stacked by rows) and
$\sigma_{t}$ be a vector of the diagonal elements of the matrix
$\Sigma_{t}$. The dynamics of the time varying
parameters is specified as random walks:
\begin{eqnarray}\label{param_model}
B_{t}=B_{t-1}+\nu_{t}, & \alpha_{t}=\alpha_{t-1}+\zeta_{t}, & \log(\sigma_{t})=\log(\sigma_{t-1})+\eta_{t},
\end{eqnarray}
where all  innovations are assumed to be jointly normally distributed and the logarithm is applied to the vector $\sigma_t$ element-wise.
In particular we assume that
\[V =
\text{Var}\left[\left(\begin{array}{c}
\epsilon_{t}\\
\nu_{t}\\
\zeta_{t}\\
\eta_{t}
\end{array}\right)\right]=\left(\begin{array}{cccc}
I_{n} & 0 & 0 & 0\\
0 & Q & 0 & 0\\
0 & 0 & G & 0\\
0 & 0 & 0 & W
\end{array}\right),
\]
where $I_{n}$ is a $n$-dimensional identity matrix, $Q,$ $G$ and
$W$ are positive definite matrices.
The prior distributions for the hyperparameters, $Q$, $W$ and the blocks of $G$,
are assumed to be  independent inverse-Wishart. The
priors for the initial states of the time varying coefficients, simultaneous
relations and log standard errors, $B_{0}$, $\alpha_{0}$ and $\log(\sigma_{0})$,
are assumed to be independent normally distributed, where the parameters of the prior distributions  are estimated by means of  the ordinary least squares (OLS) from the first $t_{0}$
observations using the regression model  \eqref{eq:exo_primiceri} (for the details see Section 4.1 of \citep{primiceri2005time}). These assumptions imply normal priors on the entire sequences of the
$B$\textquoteright s, $\alpha$\textquoteright s and log $\sigma$\textquoteright s
(conditional on $Q$, $W$ and $G$).  We use Markov Chain Monte Carlo (MCMC) technique to generate
a sample from the joint posterior of $B ,A ,\Sigma ,V$,
where $B$ is a matrix in $\mathbb{R}^{kn \times (T-t_0)}$, which 
 contains the path of the coefficients $B_{t},$  $A$ contains $a_t$,  and $\Sigma$ contains $\sigma_t$ for $t=t_{0}+1,\dots,T$.
In particular, Gibbs sampling \citep{carter1994gibbs}  is used in order to exploit
the blocking structure of the unknowns. Gibbs sampling is carried
out in four steps, returning draws of the time varying coefficients $(B )$,
simultaneous relations $(A )$, volatilities $(\Sigma )$
and hyperparameters $(V)$, conditional on the observed data and the
rest of the parameters. Conditional on $A $ and $\Sigma $, the inference for the state space model defined by \eqref{eq:primiceri} and \eqref{param_model} is carried out with the help of the Kalman filter \citep{hamilton1995time}.
The conditional posterior of $B $ is a product of
Gaussian densities, therefore $B $ can be sampled using a standard simulation
smoother \citep{carter1994gibbs}. 
For the same reason, the posterior distribution  of $A ,$ conditionally
on $B $ and $\Sigma ,$ is also a product of normal distributions.
Hence $A $ can be drawn in the same way. Remind that the process $A_t u_t$ is the product of $\Sigma_t$ and $e_t$, which is a nonlinear system of measurement equations (see equation \eqref{vol_process}). This system can be transformed into a non-Gaussian state space model  by squaring and taking logarithms for every $t$:
$$
2\log (A_t u_t) = 2\log(\sigma_t)+\log(e_t^2),
$$
where $\log(\sigma_t)$ is a random walk (\ref{param_model}).
Despite being linear, this system has innovations $\log(e_t^2)$ distributed as $\log \chi^2(1)$. We approximate the system  \textcolor{black}{with the help of a mixture of Gaussians following  \citep{primiceri2005time,kim1998stochastic,carter1994gibbs}. }
We adopt this scheme for estimation in the model \eqref{eq:exo_primiceri} under the constraint \eqref{eq:constraint}.
With a slight abuse of notations we model parameters  $B_t$, $A_t$, $\Sigma_t$ of  \eqref{eq:exo_primiceri} in the same way as described above for \eqref{eq:primiceri}.
Without the constraint \eqref{eq:constraint} the extension of the model \eqref{eq:primiceri}  to the exogenous observations is straightforward if one assumes 
\begin{equation}
D_{i,t}=D_{i,t-1}+\bar{\nu}_{i,t}, 
\label{exo_params}
\end{equation}
where the innovations \((\bar{\nu}_{i,t})\) are jointly normally distributed and independent of $\eta_t$, $\nu_t$, $\zeta_t$.  Imposing  multiplier  constraints \eqref{eq:primiceri} introduces a relation between $D_{i,t}$, $i=0,\dots,k$, which allows us to express one of the coefficients as
$$
D_{0,t} =\left[I_{n}-\sum_{j=1}^{k}B_{j,t}\right] \theta  - \sum_{i=1}^{k}D_{i,t}.
$$
We estimate parameters of the prior distributions for $B_0$,  $D_{0,i}$, $i=0,\dots,k$, $\log (\sigma_0)$ from the first $t_0$ observations using OLS ( see  \eqref{eq:exo_primiceri}).
Denote 
\[\tilde{V} =\left(\begin{array}{cc}
V & 0 \\
0 & \tilde{Q}
\end{array}\right),
\]
where $\tilde{Q}$ is a covariance of $\bar{\nu}_{i,t}$, $i=1,\dots,k$ with independent inverse-Wishart prior.
We assume a prior distribution for multiplier $\theta$  to be Gaussian $\mathcal{N}(\mu_0, U_{0})$, where the parameter  $\mu_0 \in \mathbb{R}^n$ is  estimated from a relation  \eqref{eq:constraint} for $\theta$ from estimates of  $B_0$,  $D_{0,i}$, $i=0,\dots,k$. 


The covariance matrix $U_0\in\mathbb{R}^{n\times n}$ is a diagonal matrix with large diagonal elements (uninformative prior). We propose a Gibbs sampling scheme to generate sample paths from the joint posterior of $B ,A ,\Sigma $ as well as to estimate $\theta$. The details are given in the next section for an example of modeling  gross domestic product (GDP) and real effective exchange rate (ER) with exogenous oil price. We demonstrate the performance of the proposed method for  the case of Russian and Norwegian economies.

\section{Modeling of gross domestic product and real effective exchange}
\label{sec:ru_model}
The main goal of the following setup is to model the gross domestic product (GDP) and the real effective exchange
rate (ER) while treating oil price as an exogenous variable under a long-run constraint.
Let $y_{t}\in\mathbb{R}^{2}$
be an endogenous vector, whose  first component $y_{1,t}$ denotes the difference of the logarithms of real
effective exchange rate $S_{1,t}$, the second component $y_{2,t}$ stands 
for the difference of logarithms of GDP $S_{2,t}$:
$y_{i,t}=\log (S_{i,t}/S_{i,t-1}).$
Denote the difference of the logarithms of the exogenous oil price  $S_{x,t}$ at time $t$ by
$
x_{t}=\log\frac{S_{x,t}}{S_{x,t-1}}.
$
We use \eqref{eq:exo_primiceri} to model $y_t$:
\begin{equation}
y_{t}=c_{t}+\beta_{t} y_{t-1}+D_{0,t}x_{t}+D_{1,t}x_{t-1}+u_{t},\quad t=1,\dots,T,\label{eq:gdp_exo-1}
\end{equation}
were $c_{t}\in\mathbb{R}^{2}$, $\beta_{t}=\begin{bmatrix}B_{11}(t) & B_{12}(t)\\
B_{21}(t) & B_{22}(t)
\end{bmatrix}$ , $D_{i,t}=\begin{bmatrix}D_{i,t}^{1}\\
D_{i,t}^{2}
\end{bmatrix}$, $i\in\left\{ 0,1\right\}$, $u_{t}$ is independent from $x_{t}.$
Under the constraint \eqref{eq:constraint} we have
\begin{equation}
\theta=\left[I_{n}-\beta_{t}\right]^{-1}(D_{0,t}+D_{1,t}),\label{eq:constraint-2}
\end{equation}
where $\theta\in\mathbb{R}^{2}$ is an unobserved multiplier parameter. As our variables enter the model  logarithmically, the parameter $\theta$ has an interpretation of the long-run elasticity.
From \eqref{eq:constraint-2} we derive
\[
D_{0,t}=\left[I_{n}-\beta_{t}\right]\theta-D_{1,t}
\]
and therefore for a fixed $\theta$ the corresponding measurement equation reads as follows
\begin{equation}
y_{t}-\theta x_{t}=c_{t}+\beta_{t}[y_{t-1}-\theta x_{t}] -{D}_{1,t}(x_{t}-x_{t-1})+u_{t},\,t=1,\dots,T.\label{eq:elast_inserted}
\end{equation}

\subsection{Gibbs sampling with elasticity estimation}

First we use OLS to estimate parameters (means and variances) of the prior distributions for  $B_0$,  $D_{i,0}$, $i=0,1$ 
for initial states of parameters $B_{t}$ (vectorized $\beta_t$), $D_{0,t},$ $D_{1,t}$ without
elasticity constraints. 
We assume a prior  distribution for $\theta$  to be  $\mathcal{N}(\mu_0, U_{0})$, where the vector $\mu_0 \in \mathbb{R}^n$ is  estimated   from \eqref{eq:constraint-2}  given estimates of the mean of prior distributions of $B_0$,  $D_{i,0}$, $i=0,1$. The entries of the diagonal matrix $U_0$ have large values. 
Denote by  $s$  a vector with  indicator variables in Gaussian mixture approximation which takes part in estimating $u_t$ (see Section \ref{sec:Bayesian-inference-with}). Denote the  trajectories of all  parameters for a fixed value of $\theta^{(j-1)}$ at $j$th MCMC simulation step by 
$$Z^{(j)}_{\theta} = [B^{(j)},  {D}^{(j)}_1, \Sigma^{(j)}, A^{(j)}, \tilde{V}^{(j)}].$$
The steps of the proposed Gibbs sampling scheme are as follows. 
\begin{enumerate}
\item Draw $Z^{(j)}_{\theta}$ conditionally on $\theta^{(j-1)}$ and $Y$ based on the model \eqref{eq:elast_inserted}. Denote by $p$ and $\tilde{p}$ the likelihood and the approximated likelihood (using Gaussian mixtures), respectively,  and $\upsilon =[B, {D_1},A,\tilde{V}].$  We proceed with the following sampling steps  (see \citep{del2015time} for details)  by drawing:
\begin{itemize}
\item $\Sigma$ from $\tilde{p}(\Sigma|Y,\upsilon ,s)$,
\item $\upsilon$ from $p(\upsilon|Y,\Sigma)$,
\item $s$ from $\tilde{p}(s|Y,\Sigma,\upsilon).$
\end{itemize}

\item Draw $\theta^{(j)}$ conditionally on $Z^{(j)}_{\theta}$ from
$
\mathcal{N}(\mu_{j},U_{j}).
$
The parameters  of the posterior distribution $\mu_{j}$, $v^2_{j}$ are estimated 
 from the observations
\[
\tilde{Y_{t}}^{(j)}=C_{t}^{(j)}\theta+u_{t}^{(j)},\quad t=t_0+1,\dots,T,
\]
where
$\tilde{{Y}}_{t}^{(j)}=y_{t}-c_{t}- \beta_{t}^{(j)} y_{t-1}- {D}_{1,t}^{(j)}(x_{t-1}-x_{t}),$
$C_{t}^{(j)}=x_{t} [I-\beta_{t}^{(j)}] $
and $u_{t}^{(j)}\sim\mathcal{N}(0,H_{t}^{(j)}).$
Therefore the posterior distribution of $\theta^{(j)}$ is defined by the covariance
$$
U_{j}^{-1}=U_{0}^{-1}+\sum_{t=t_0+1}^{T}C_{t}^{(j)\top}[H_{t}^{(j)}]^{-1}C_{t}^{(j)},
$$
and by the mean
$$
\mu_{j}=U_j\left(U_{0}^{-1}\theta_{0}+\sum_{t=t_0+1}^{T}C_{t}^{(j)\top}[H_{t}^{(j)}]^{-1}\tilde{Y_{t}}^{(j)}\right).
$$
\end{enumerate}

\subsection{Impulse-response analysis}

Impulse response characterization demonstrates the behavior of the output after a small shock in the input variable.
We shall be  interested in $10\%$ increase of oil prices obtained by a single shock $x_t=\log(1.1)$ in the model \eqref{eq:gdp_exo-1} with
\eqref{eq:constraint-2}.
The  shock evolves according to \eqref{eq:gdp_exo-1}  as
\begin{eqnarray*}
\delta y_{t+1} &= &\log(1.1)D_{0,t},\\
\delta  y_{t+2} &=&\log(1.1)[\beta_{t}  D_{0,t}+D_{1,t}],\\
\ldots\\
\delta  y_{t+k} &=&\log(1.1)[\beta_{t} ^k D_{0,t}+\beta_{t} ^{k-1}D_{1,t}].
\end{eqnarray*}
Hence, a change in the logarithm (element-wise) of  the vector $S_t$ reads as
$$
\log  {S_{t+k}}-\log{S_t} = \sum_{j=1}^k \delta  y_{t+j}.
$$
Thus, when $k\to \infty$ we get
\begin{equation}\label{irf_limit}
\log  {S_{t+k}}-\log{S_t}  \to [I-\beta_{t}]^{-1}(D_{1,t}+D_{0,t})\log(1.1)=\log(1.1)\theta.
\end{equation}
Therefore $\theta$ defines a fully adjusted value of a  response after a shock and  describes the underlying permanent state of economy.

\section{Numerical results}

This section contains empirical analysis of the proposed constrained TVP-VAR method based on economic data for Norwegian and  Russian economies. We have selected these countries for the analysis because they are among the top oil-exporters. Furthermore both Russia and Norway underwent significant changes in exchange rate policy in historical retrospective. 
We compare performance of the constrained TVP-VAR for modeling the Norwegian and the Russian economies with the following benchmark methods: 1) method from \citep{del2015time,primiceri2005time} extended for estimation of the model with exogenous variables \eqref{eq:gdp_exo-1} with no elasticity restrictions \footnote{The code for the proposed method and the first benchmark method is based on modifications of a CRAN package \citep{bvarsv}}, 2) VAR with constant parameters. For  comparison  we computed the absolute value of the deviation of out-of-sample forecasts  from the corresponding observations of GDP and real exchange rate for the number of steps ahead lying in the set $\{1,2,3,4,5\}$.
\subsection{Norwegian economy}
For Norway we had $157$ quarterly observations of the real exchange rate $S_{1,t}$,
GDP $S_{2,t}$, and oil price $S_{x,t}$ staring from the 1st quarter  1980  till 1st quarter 2019. Therefore  the number of logarithm
differences $y_t$ and $x_t$  of observations $S_{x,t}$  and $S_{i,t}$, $i=1,2$ was $156$. %
%
For the estimation of prior parameters of the constrained TVP-VAR
we used the first $t_{0}=40$ observations of $y_{i}^t$, $i=1,2$ and $x_t$.  We use an uninformative prior for the elasticity  $\theta\sim\mathcal{N}(\mu_0, U_{0})$ (see Section \ref{sec:Bayesian-inference-with}) with $ U_{0}={\rm diag}([0.1,0.1])$.

\begin{figure}[h]
	\begin{center}
		\includegraphics[scale=0.55]{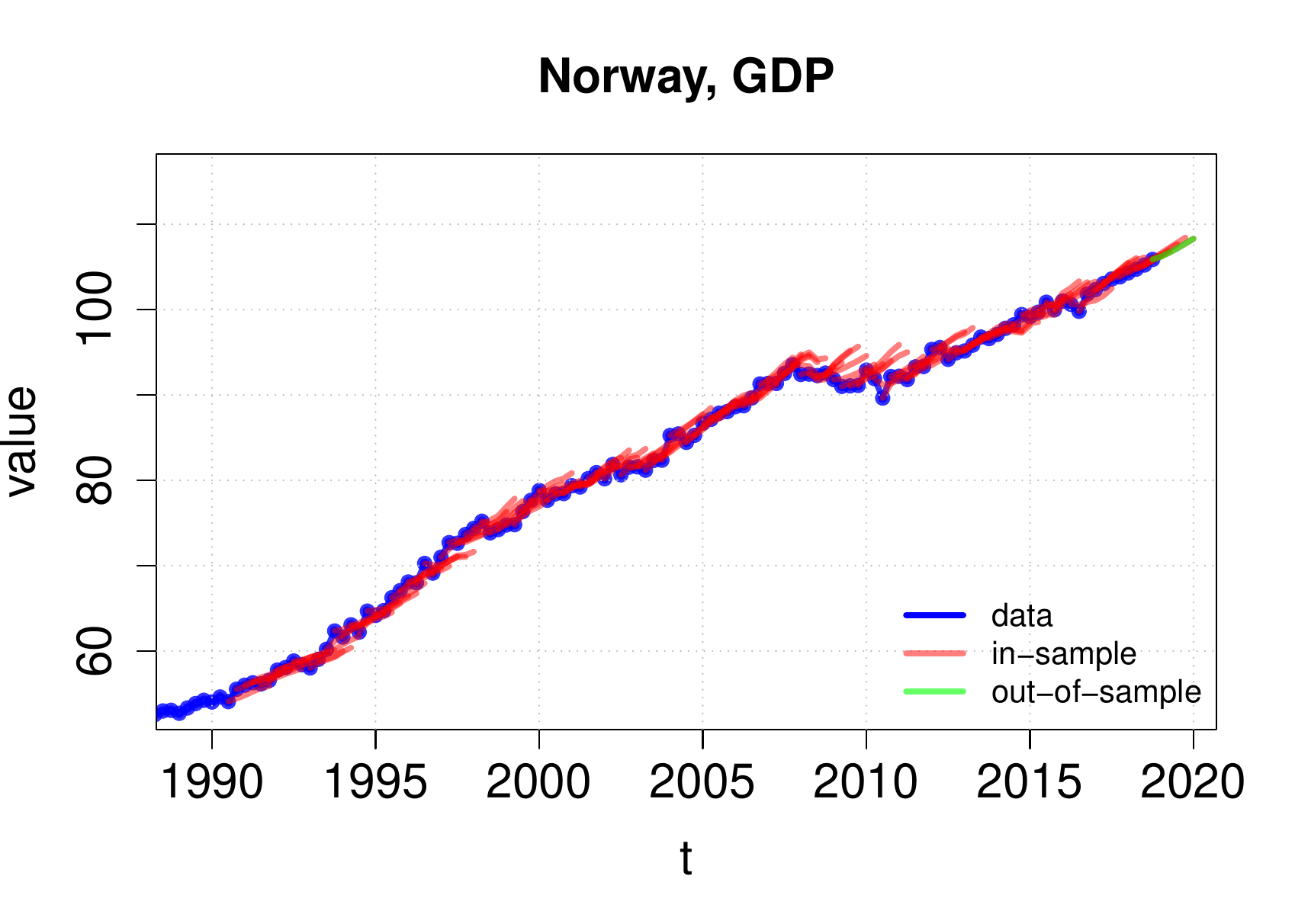}
		\caption{\label{fig:Forecasts_No_gdp}Five steps ahead in-sample (red) and out-of sample (green) forecasts for Norwegian  GDP by the model \eqref{eq:gdp_exo-1} with
			elasticity constraint \eqref{eq:constraint-2}}
				\end{center}
	\end{figure}

\begin{figure}[h]
	\begin{center}
		\includegraphics[scale=0.55]{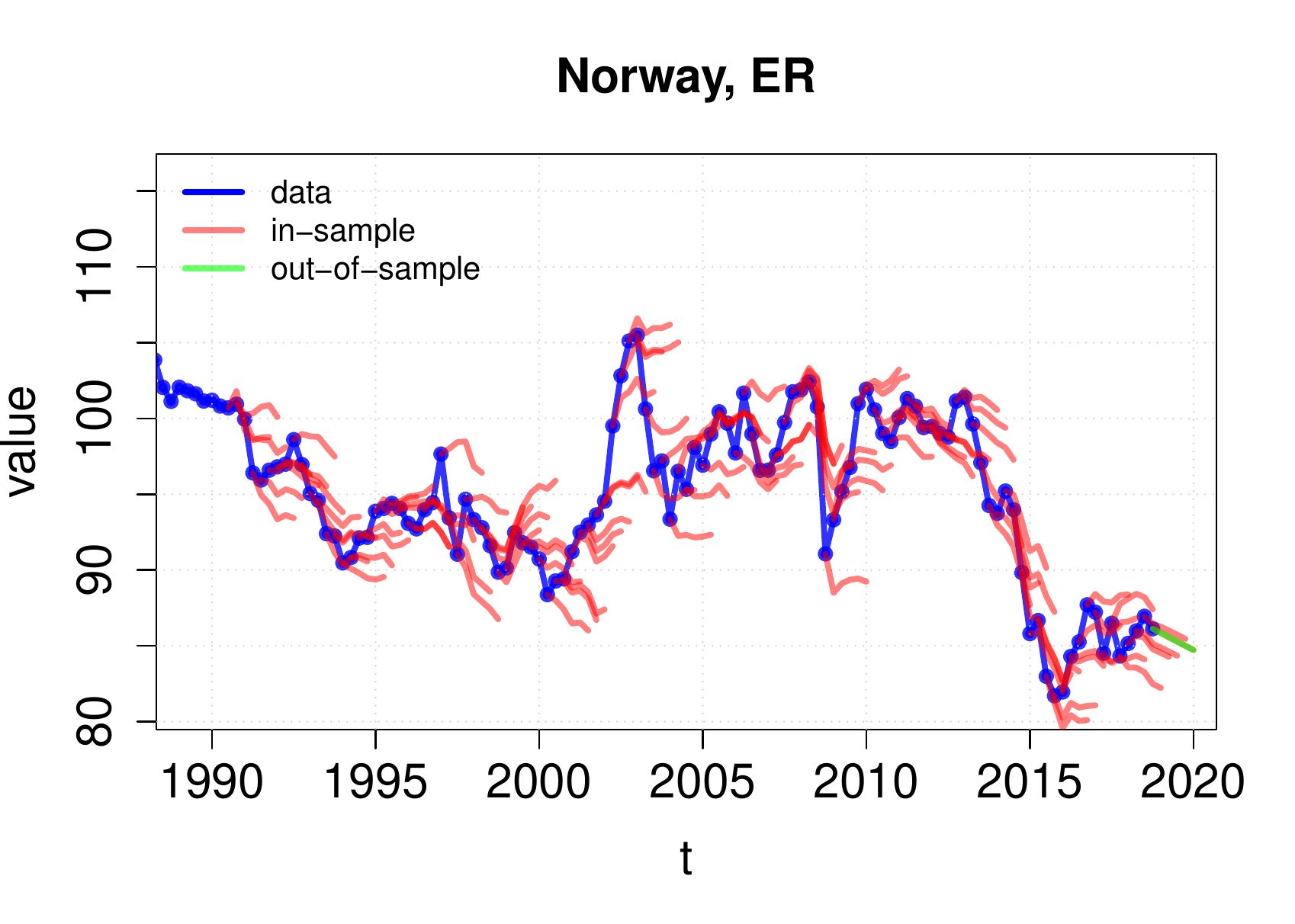}
		\caption{\label{fig:Forecasts_No_er}Five steps ahead in-sample (red) and out-of sample (green) forecasts for Norwegian real effective exchange rate by the model \eqref{eq:gdp_exo-1} with
			elasticity constraint \eqref{eq:constraint-2}}
	\end{center}
\end{figure}

\begin{figure}[h]
	
	\includegraphics[scale=0.4]{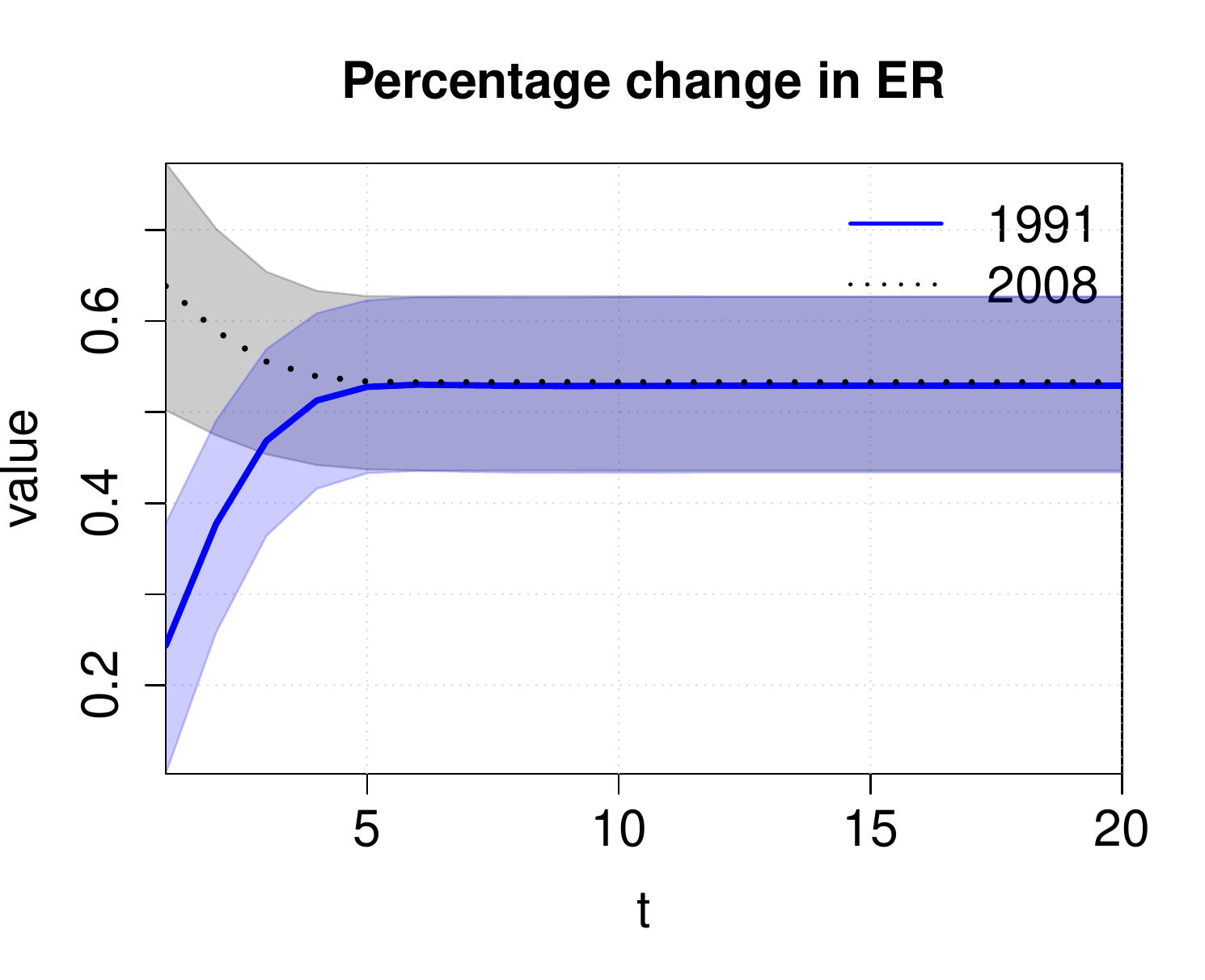}
	\includegraphics[scale=0.4]{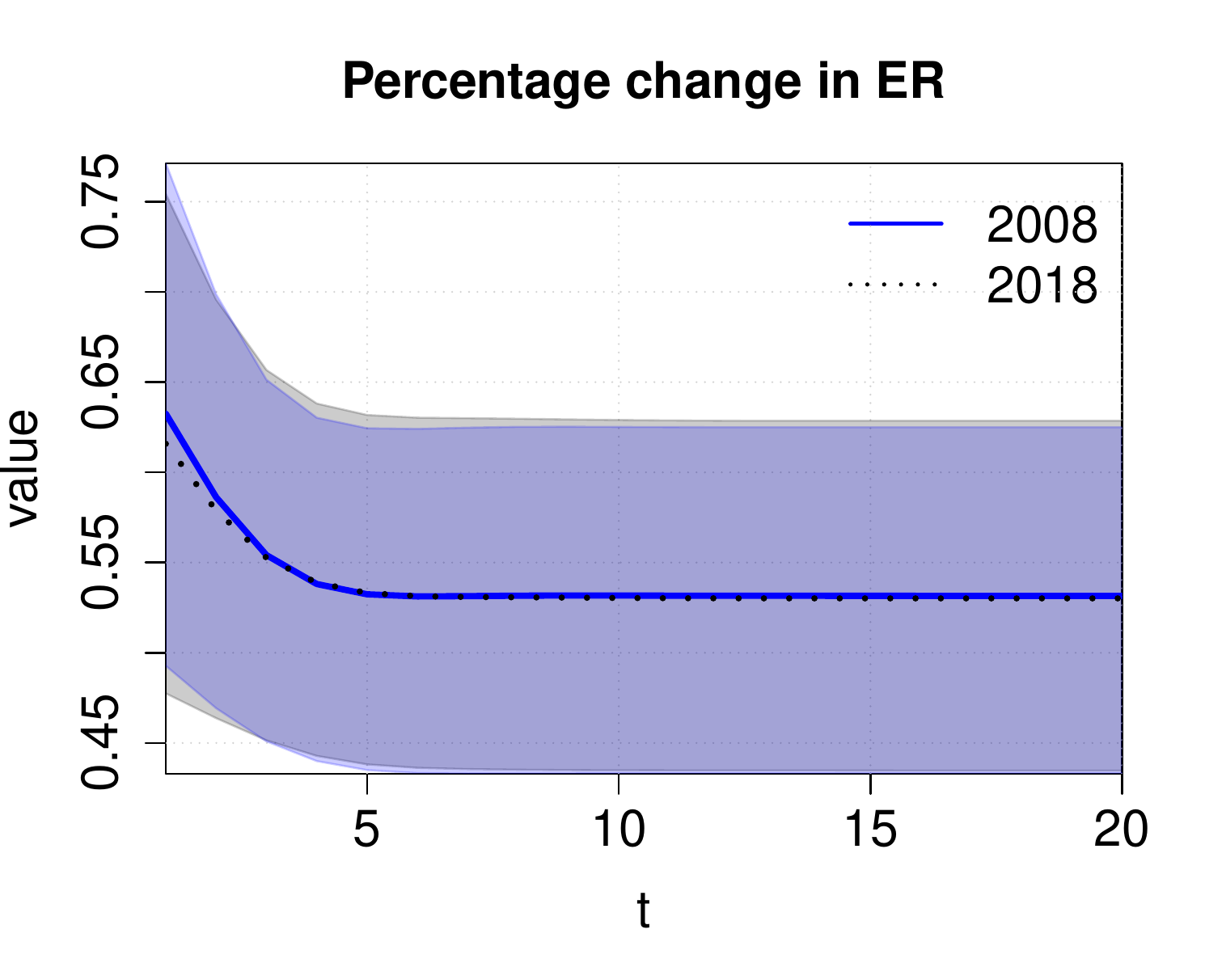}
	\caption{\label{fig:IRF_ER_No} Norway.  Impulse response functions for the proposed model:  real exchange rate as a response variable}  
	\includegraphics[scale=0.4]{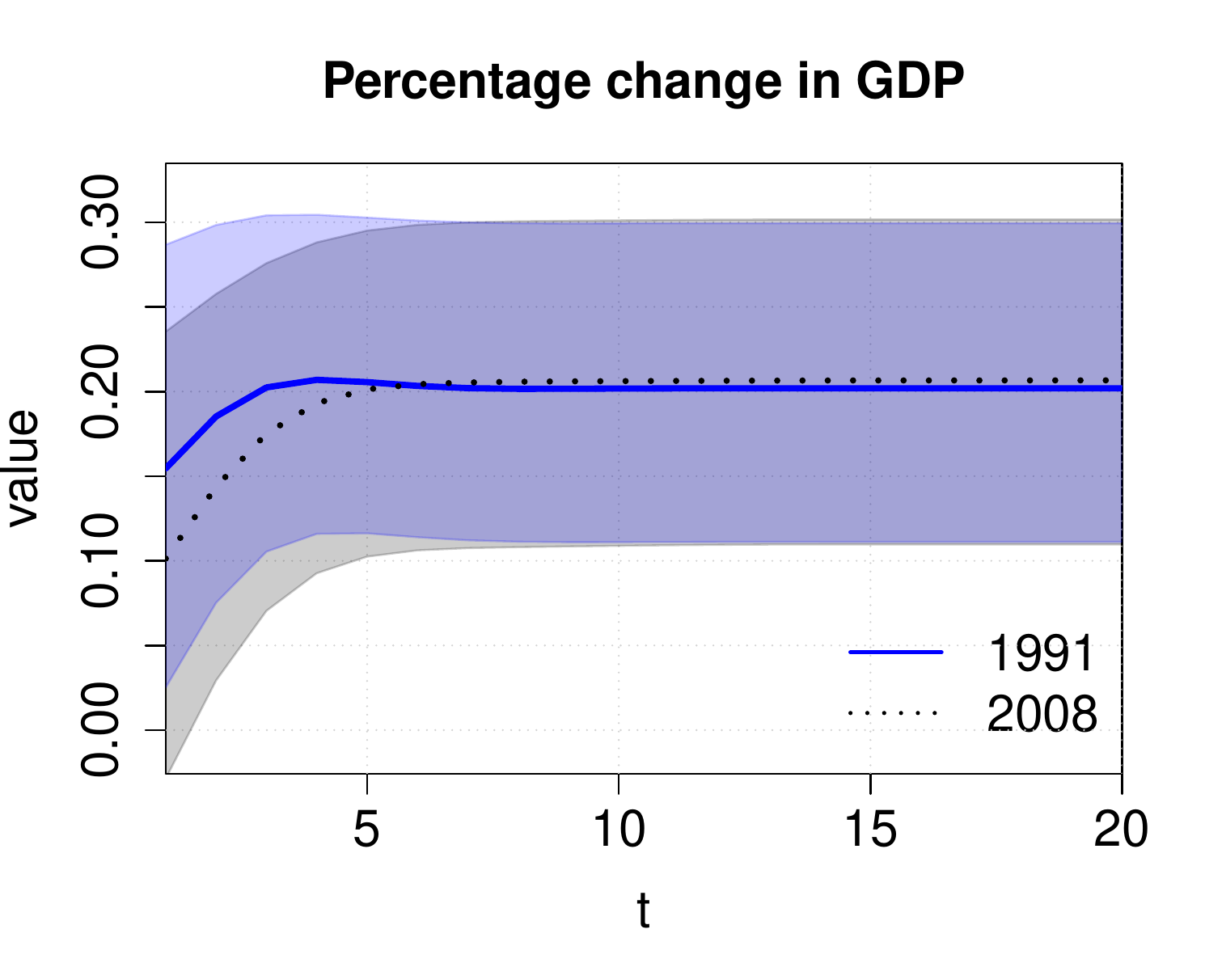}
	\includegraphics[scale=0.4]{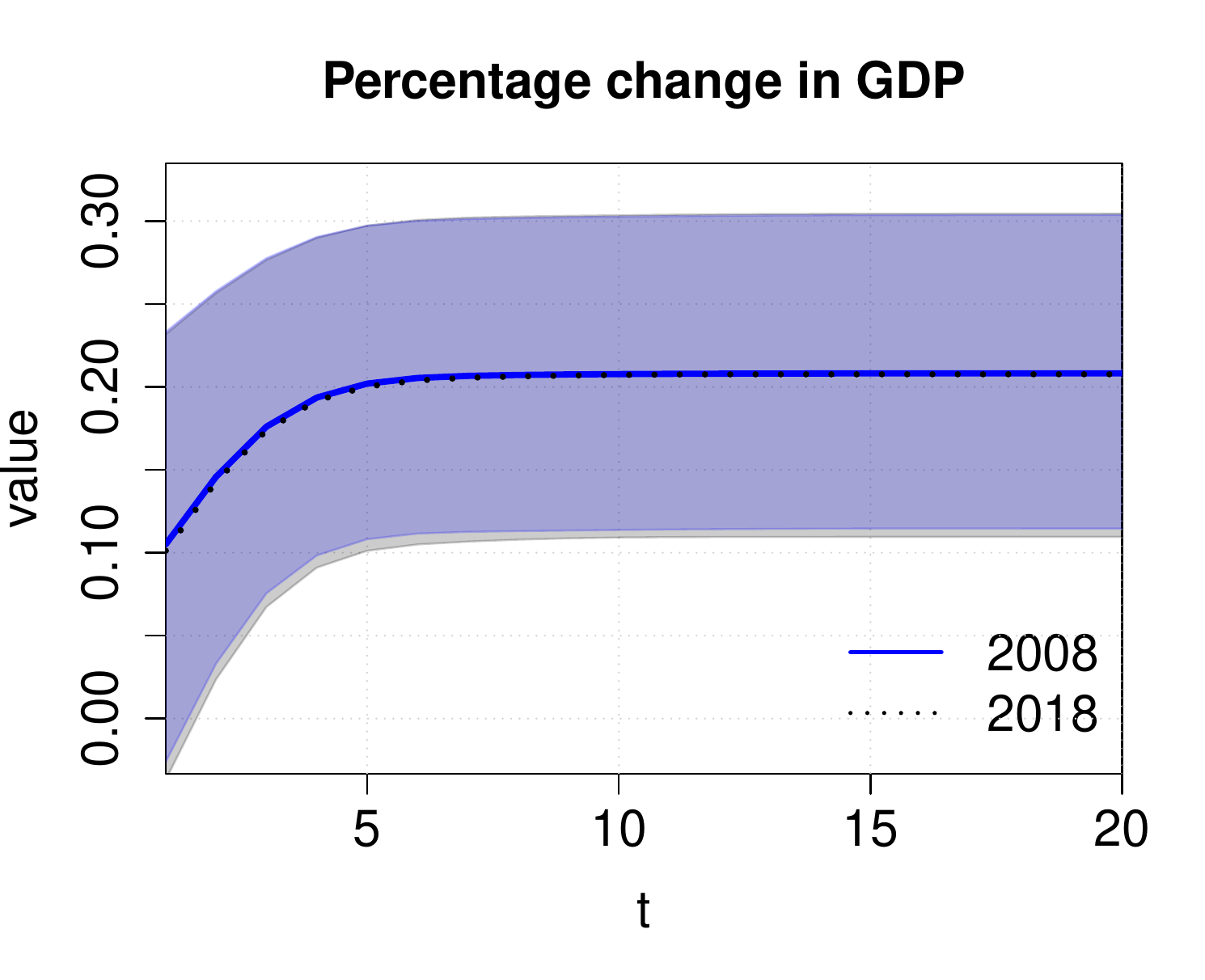}
	\caption{\label{fig:IRF_GDP_No}Norway. Impulse response functions for the proposed model:  GDP  as a response variable } 
	\end{figure}

After $30000$ MCMC steps the estimation procedure converged to $\theta=[0.05,\, 0.02]^{\top}$, where the first component corresponds to real exchange rate and the second to GDP.  In-sample forecasts by the constrained TVP-VAR for the time interval $[1992\text{Q}_4,2019\text{Q}_1]$ are shown in Figures \ref{fig:Forecasts_No_gdp}, \ref{fig:Forecasts_No_er}. Figures \ref{fig:IRF_ER_No}, \ref{fig:IRF_GDP_No} contain IRF for years 1991,  2008, 2018 whereas Fig.\ref{fig:IRF_3D_No} contains 3D IRF for GDP and real exchange rate.

The errors of 1-5 step ahead out-of-sample forecasts for  the proposed method  and benchmark methods for the time interval $[1992\text{Q}_4,2019\text{Q}_1]$ are collected  in Table \ref{error_an_no}.
Long-run  the  impulse responses  to a positive shock in oil prices  are positive for both the real exchange rate and  real GDP.
Improvement in the terms of trade leads to the exchange rate strengthening, which ensures internal and external equilibrium.
This means that  for the same volume of exports a country can buy a larger volume of imported goods.  Therefore the prices of domestic non-tradable goods relative to prices of imported goods should increase to ensure the increase in the share of imported goods in aggregated consumption \citep{edwards1988real}.
Furthermore the oil prices rise leads to an increase in GDP through the  capital accumulation channel, namely, the higher oil prices result in new  investment opportunities  \citep{esfahani2014empirical} and  increase of domestic returns \citep{idrisov2015theoretical}.
The model indicates significant change in the short run transmission mechanism of oil prices shocks to the real exchange rate. IRFs for years 1991 and 2018 are statistically different. Before the Norges Bank turned to inflation targeting in 2001 the real exchange rate response had been strengthening gradually towards its long-run equilibrium after a shock in oil prices.
Under the inflation targeting regime we see some overshooting of the real exchange rate.
There is no sizable time variation in model parameters over  the last decade, and the shape of the impulse response function for the real exchange rate stabilizes. Impulse response function for the real GDP changes very slightly during the entire period under review.
The results of pseudo out-of-sample forecasting experiment in Table \ref{error_an_no} show that the proposed TVP-VAR with time-invariant long-run multipliers outperforms the benchmark TVP-VAR without constraints. Thus reduction of degrees of freedom in TVP-VAR model helps to improve the forecasting accuracy.
The proposed constrained TVP-VAR delivers smaller forecast errors than constant-parameter VAR  for 1-3 steps-ahead forecasts and  accuracy similar to  VAR for 4-5 step-ahead forecasts.

Next we consider an example of Russian economy, 
were a transition towards the  inflation targeting regime has started in  2014.

\begin{figure}[h]
	\begin{center}
		\includegraphics[scale=0.35]{Norway_IRF_sur_GDP.pdf}
		\includegraphics[scale=0.35]{Norway_IRF_sur_ER.pdf}
	\end{center}
	\caption{\label{fig:IRF_3D_No} Norway. 3D impulse response functions for the proposed model: the impulse
		in oil prices and GDP (left) and real effective exchange rate (right) as response variables correspondingly}  
\end{figure}

\FloatBarrier
\begin{table}[H]
	\centering
	\ra{1.3}
	\begin{tabular}{@{}rrrcrrcrr@{}}\toprule
		& \multicolumn{2}{c}{Constrained} & \phantom{abc}& \multicolumn{2}{c}{VAR} &
		\phantom{abc} & \multicolumn{2}{c}{TVP-VAR \eqref{eq:gdp_exo-1}}\\ \cmidrule{2-3} \cmidrule{5-6} \cmidrule{8-9}
		steps & mean & std &&  mean & std &&  mean & std \\ \midrule
		GDP    \\
1 & 0.60 & 0.61 && 1.02 & 0.79 && 1.02 & 0.80 \\
 2 & 0.95 & 0.84 && 1.07 & 0.89 && 1.08 & 0.90 \\
 3 & 1.16 & 1.04 && 1.26 & 1.01 && 1.26 & 0.97 \\
 4 & 1.39 & 1.25 && 1.45 & 1.15 && 1.40 & 1.12 \\
 5 & 1.62 & 1.39 && 1.61 & 1.27 && 1.55 & 1.26 \\
		ER  \\
1 & 1.31 & 1.24 && 1.39 & 1.17 && 1.57 & 1.34 \\
 2 & 1.97 & 1.65 & & 2.21 & 1.68 && 2.53 & 1.88 \\
 3 & 2.71 & 2.04 && 2.75 & 2.10 && 3.11 & 2.16 \\
 4 & 3.18 & 2.51 && 3.07 & 2.60 && 3.43 & 2.64 \\
 5 & 3.33 & 2.80 && 3.28 & 2.75 && 3.68 & 2.76 \\ \bottomrule
	\end{tabular}
	\caption{\label{error_an_no} Norway. Mean and standard deviation of absolute error of out-of-sample forecasts for the proposed method, TVP-VAR \eqref{eq:gdp_exo-1} with exogenous variables and VAR.}
\end{table}

\FloatBarrier
\subsection{Russian economy}

We use $93$ quarterly observations of real effective exchange rate of  $S_{1,t}$,
GDP $S_{2,t}$ for Russia and oil price $S_{x,t}$ from the 1st quarter 1995 till the 4th quarter 2018. A number of logarithm
differences of observations of $y_t$ therefore was $92$.
%
%
For the estimation of prior parameters of constrained TVP-VAR
we used the first $t_{0}=40$ observations of $y_{i}^t$, $i=1,2$ and $x_t$.  We selected an uninformative Gaussian prior for the elasticity  $\theta\sim\mathcal{N}(\mu_0, U_{0})$ (see Section \ref{sec:Bayesian-inference-with}) with $ U_{0}={\rm diag}([0.1,0.1])$.

\begin{figure}[h!]
\begin{center}
\includegraphics[scale=0.5]{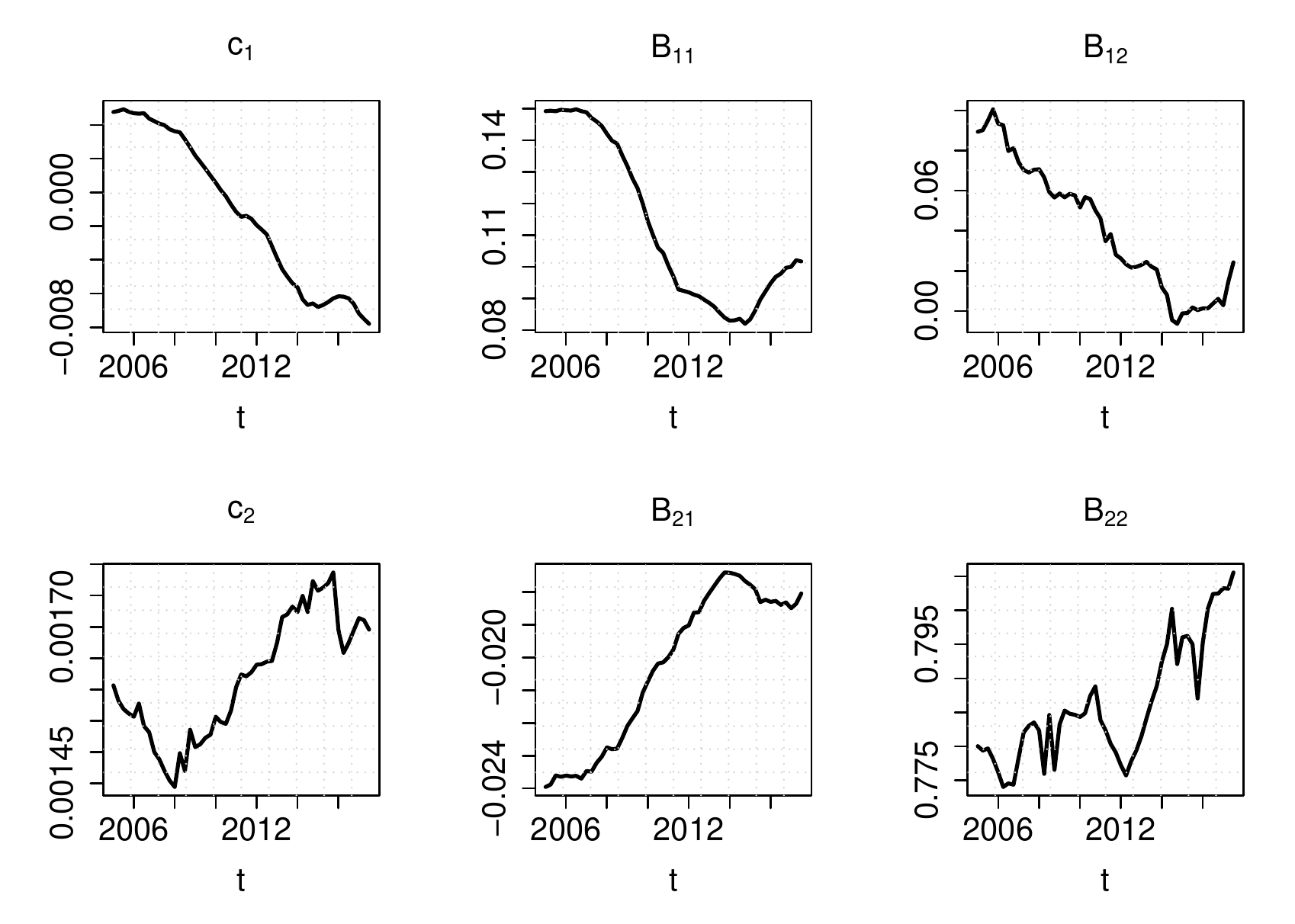}\caption{\label{fig:VAR_coefficients}VAR coefficients in \eqref{eq:gdp_exo-1} with \eqref{eq:constraint-2}: first
column corresponds to  $c_{t}$, the
next two columns show the behavior of the entries of $B_{t}$.}
\end{center}
\end{figure}

\begin{figure}[h!]
\begin{center}
\includegraphics[scale=0.50]{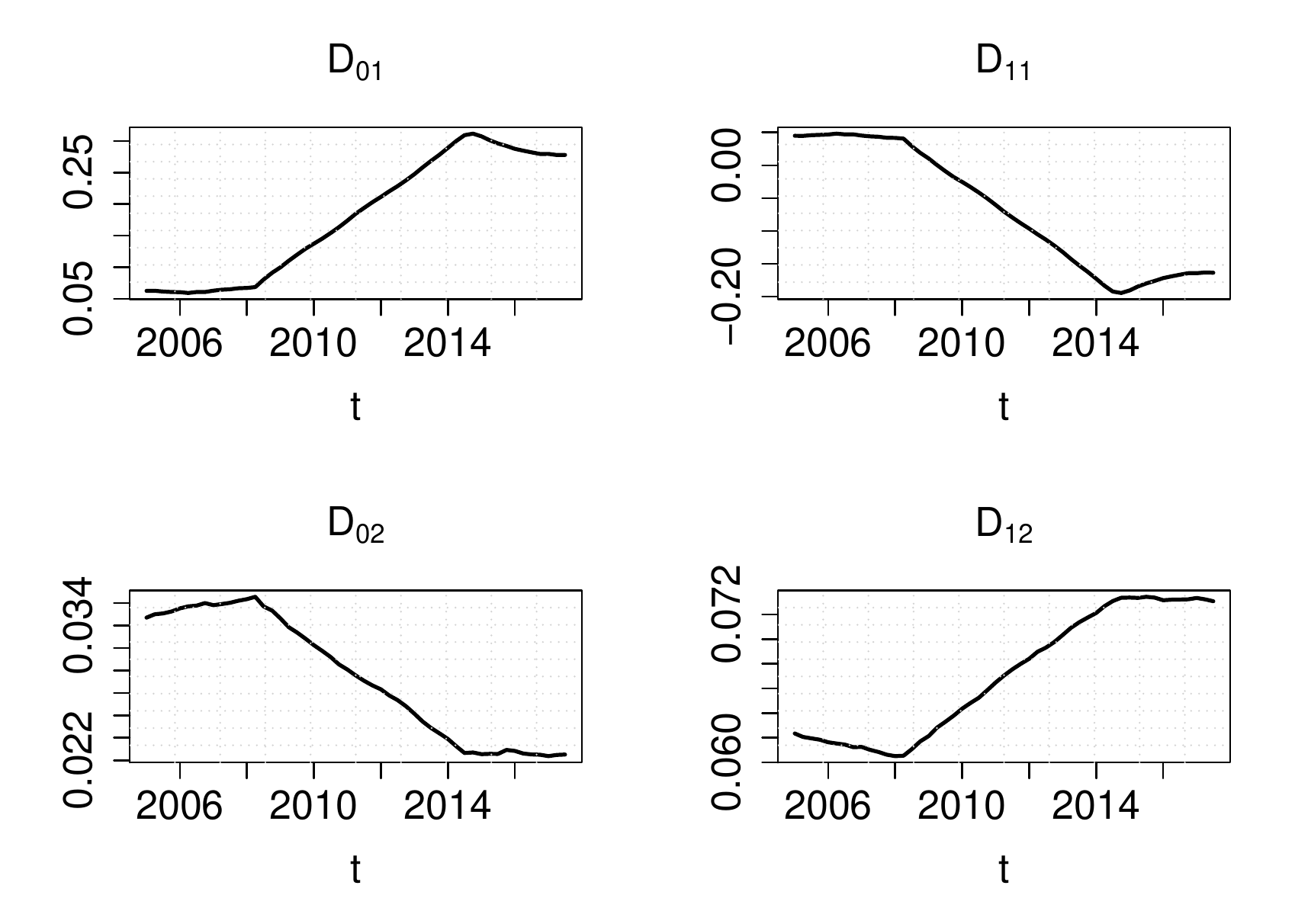}\caption{\label{exog_coeff}The dynamics of the entries of exogenous coefficients $D_{0,t}$ and $D_{1,t}.$}
\end{center}
\end{figure}

\begin{figure}[h!]
\begin{center}
\includegraphics[scale=0.580]{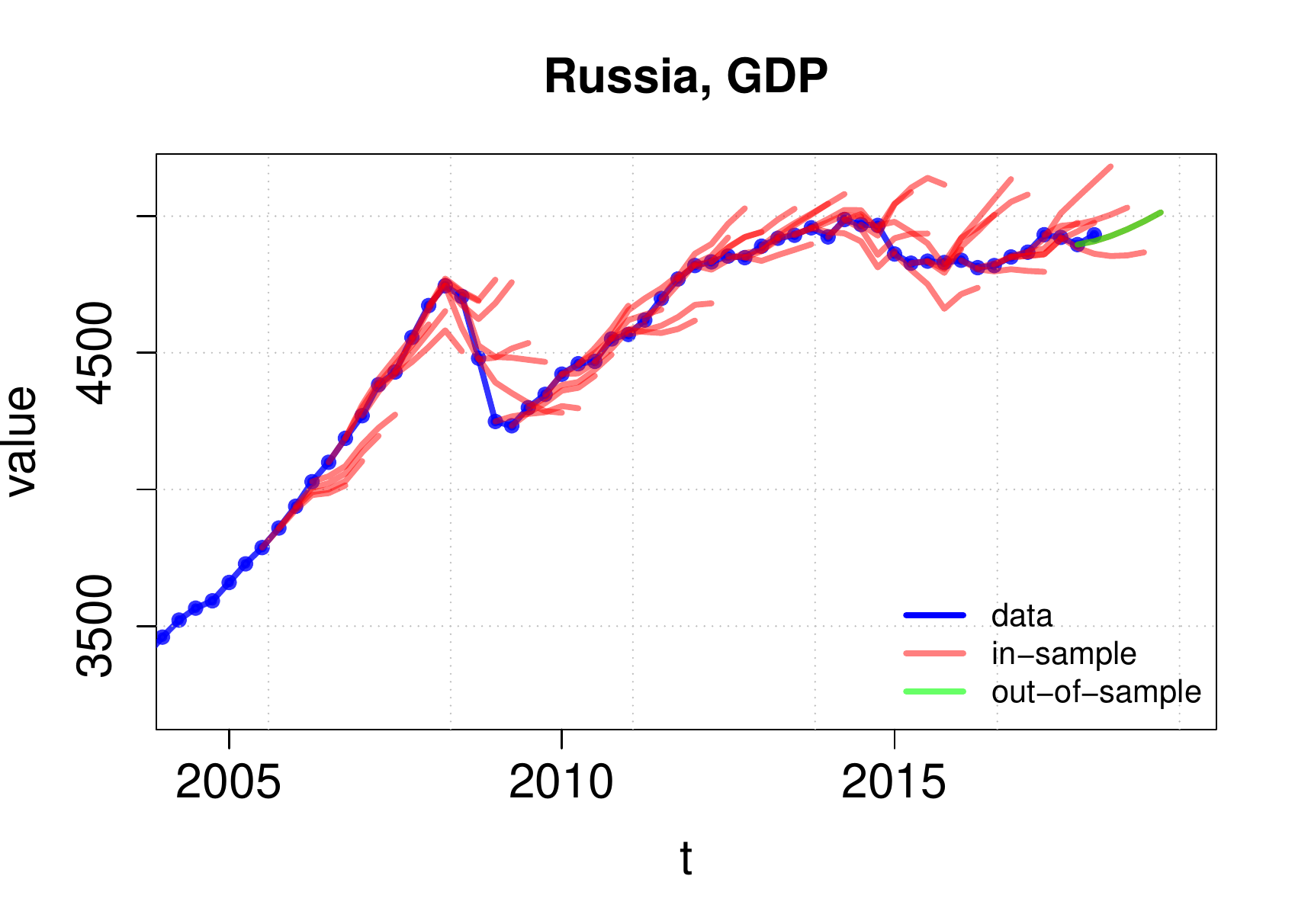}
\caption{\label{fig:Forecasts}Five steps ahead in-sample (red) and out-of sample (green)  forecasts for GDP by the model \eqref{eq:gdp_exo-1} with
elasticity constraint \eqref{eq:constraint-2}}
\end{center}
\end{figure}

\begin{figure}[h!]
\begin{center}
\includegraphics[scale=0.580]{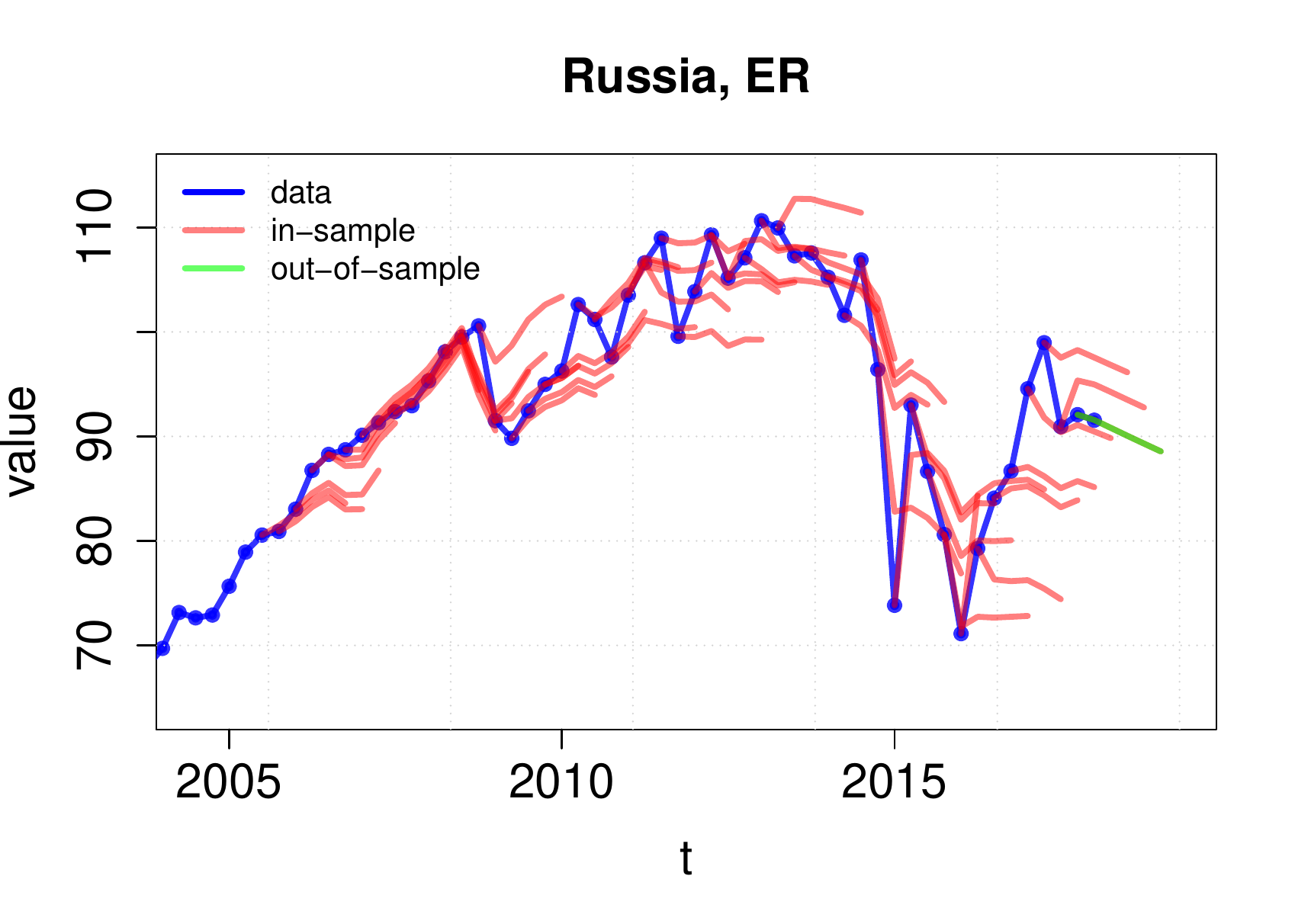}\caption{\label{fig:Forecasts-for-exchange}Five steps ahead in-sample (red) and out-of sample (green)  forecasts for the real effective
exchange rate by the model \eqref{eq:gdp_exo-1} with elasticity constraint
\eqref{eq:constraint-2}.}
\end{center}
\end{figure}

\begin{figure}[h!]
	\begin{center}
		\includegraphics[scale=0.42]{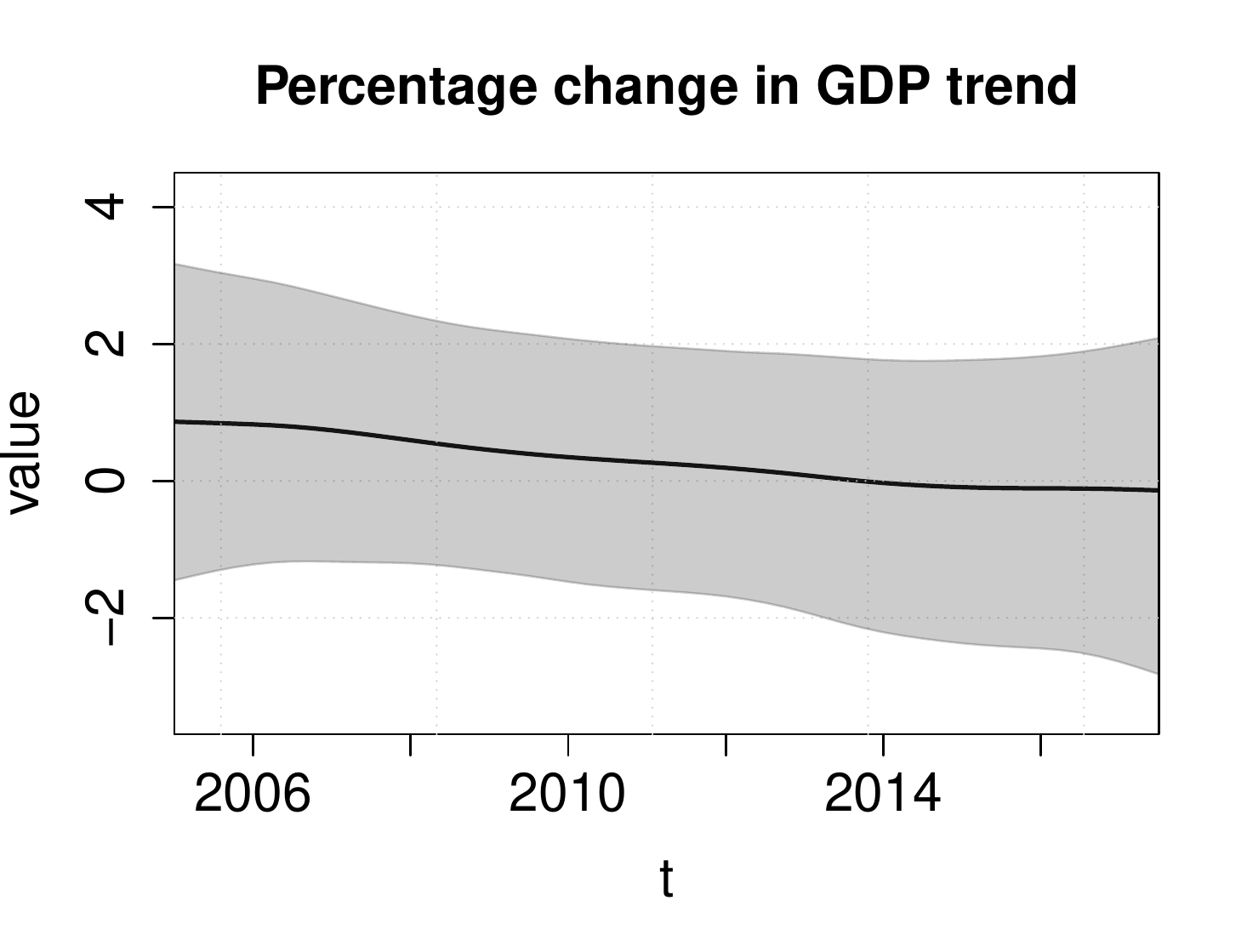}\caption{\label{fig:GDP_lr} Percents of long-run growth rate of Russian GDP with $60\%$ confidence intervals}
	\end{center}
\end{figure}

After $30000$ MCMC steps the estimating procedure for constrained TVP-VAR converged to $\theta=[0.09,\, 0.04]^{\top}$, where the first component corresponds to real exchange rate and the second to GDP.  Posterior  median of VAR part  of constrained TVP-VAR coefficients are shown in Figure  \ref{fig:VAR_coefficients},
posterior medians of $D_{0,t}$ and $D_{1,t}$  are in Figure  \ref{exog_coeff}.
The five-step ahead in-sample forecasts for the GDP and the real effective exchange
rate (ER)  along with out-of-sample forecast for the time interval  $[2007\text{Q}_4,2018\text{Q}_4]$ are  shown in Figures \ref{fig:Forecasts}, \ref{fig:Forecasts-for-exchange}. Median of long-run growth rate with $60\%$ confidence intervals for GDP, which is the second component of $(I-B_t)^{-1}c_t$, in percents is shown in Figure \ref{fig:GDP_lr}.


\begin{figure}[h!]
\includegraphics[scale=0.38]{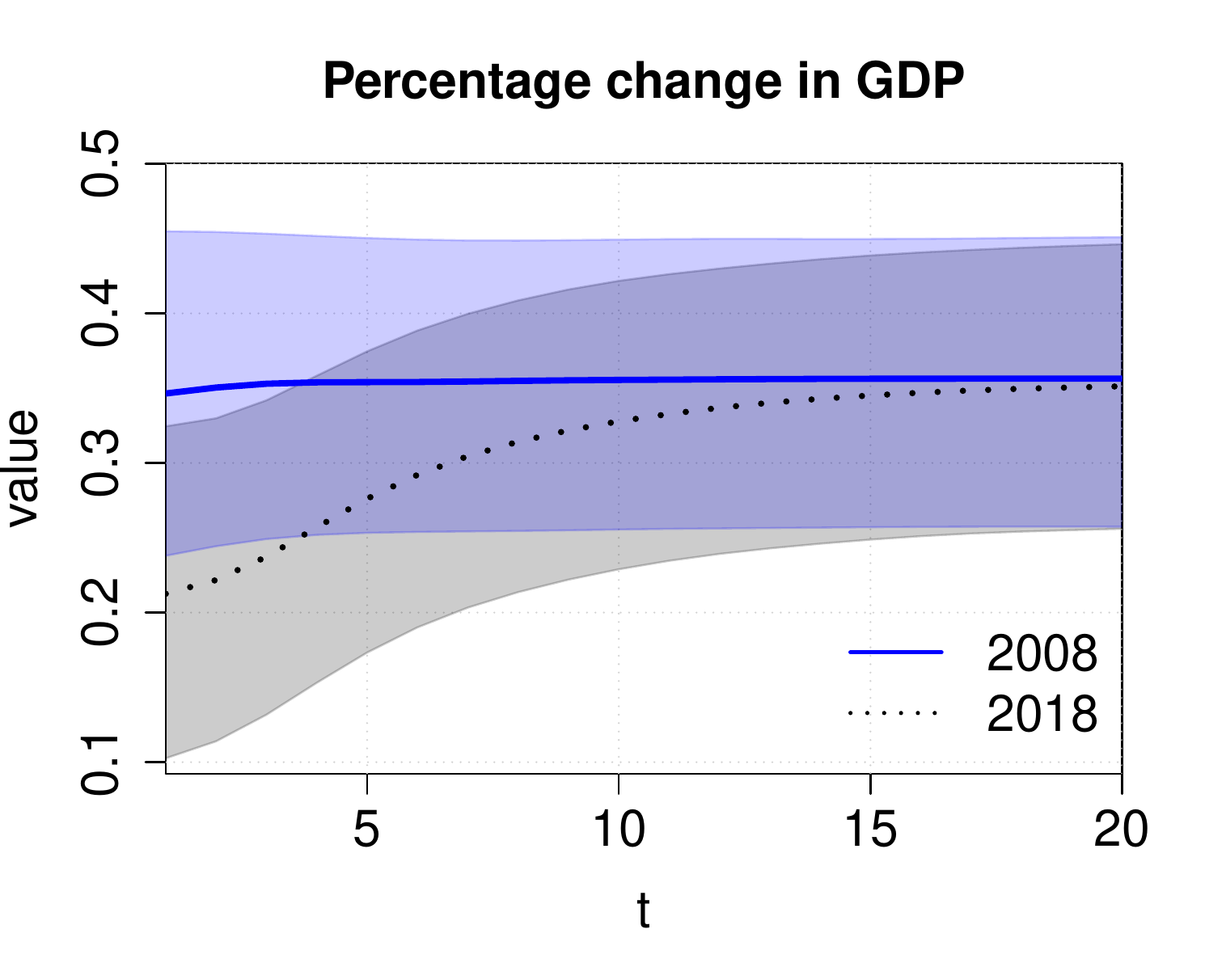}
\includegraphics[scale=0.38]{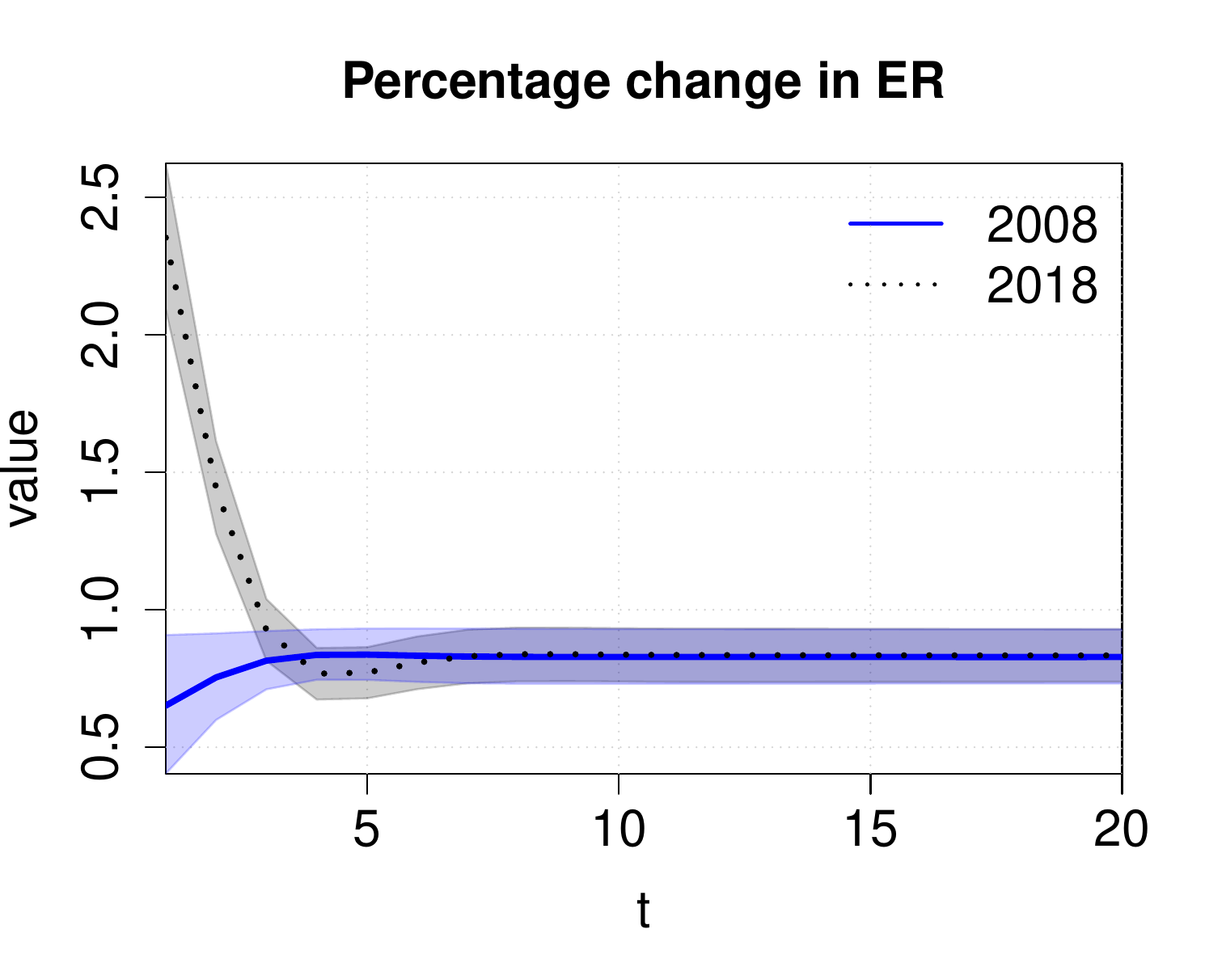}
\caption{\label{fig:IRF_GDP} Impulse response functions for the proposed model: the impulse
in oil and GDP and real exchange rate as response variable } 
\end{figure}

\begin{figure}[h!]
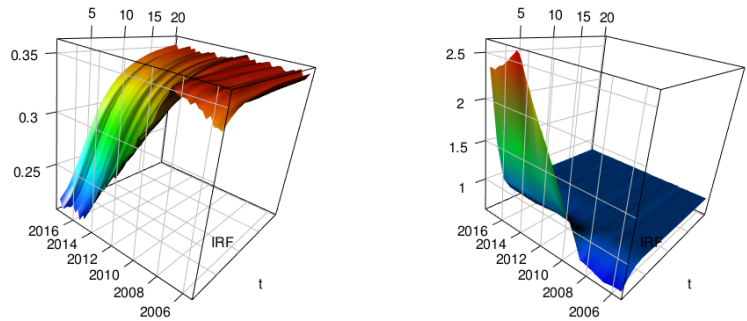

\begin{center}
\includegraphics[scale=0.33]{Ru_IRF_sur_GDP.pdf}
\includegraphics[scale=0.33]{Ru_IRF_sur_ER.pdf}
\end{center}
\caption{\label{fig:IRF_3D} 3D impulse response functions for the proposed model: the impulse
in oil and GDP (left) and real exchange rate (right) as response variables correspondingly } 
\end{figure}

We compare  impulse response functions for the years 2008 and 2018 for GDP and real exchange rate (ER)  to the shock in exogenous logarithm of differences of oil prices.
Results in Figs. \ref{fig:IRF_GDP} demonstrate the convergence to the same limiting value defined by \eqref{irf_limit}; 3D-plots of impulse response functions are shown in  Figs. \ref{fig:IRF_3D}. 
During the years  before the crisis of 2008--2009  the Central Bank of Russia followed the policy of a managed nominal ruble exchange rate. From IRF for this period one can observe a gradual strengthening of the real exchange rate  towards its long-run equilibrium after an increase in oil prices.
During the next years the Central Bank of Russia switched to a floating exchange rate.
After that the real exchange rate began to react to oil price shocks more sharply with the overshooting effect.
It should be noted that during periods of gradual reaction of the exchange rate to the oil price shocks, real GDP reacted quite strongly to the shock.
During the periods of sharp reaction of the real exchange rate the real GDP demonstrates gradual increase.
Therefore our results are in line with a classical view:  flexible exchange rates are shock absorbers for small open economies and the  floating exchange rate regime of monetary policy reduces volatility of the GDP growth.


The mean absolute errors and standard deviations for of 1-5 steps out-of-sample forecasts of GDP and real exchange rate for the proposed method and benchmark methods for the time interval $[2007\text{Q}_4,2019\text{Q}_1]$ are shown in Table  \ref{error_an}. One may conclude that in terms of forecasts a classical VAR gives better result than TVP-VAR \citep{del2015time,primiceri2005time} extended for exogenous variables case.
Imposing the elasticity constraint helps to improve the situation: the proposed method  outperforms both benchmark methods in  forecasting of GDP 1-3 steps ahead. The proposed method gives smaller forecasting error that non-constrained TVP-VAR for the real exchange rate. Nevertheless, the uncertainty coming from the coefficients model brings though delivers slightly less accuracy than VAR for the real exchange rate and 4-5 steps ahead forecasts of GDP.
The results demonstrate that imposing the long run elasticity constraint allows to improve the quality of modeling.
Figure \ref{fig:GDP_lr} demonstrates significant decrease in the long-run growth rates for the Russian economy.


\begin{table*}
\centering
\ra{1.3}
\begin{tabular}{@{}rrrcrrcrr@{}}\toprule
& \multicolumn{2}{c}{Constrained} & \phantom{abc}& \multicolumn{2}{c}{VAR} &
\phantom{abc} & \multicolumn{2}{c}{TVP-VAR \eqref{eq:gdp_exo-1}}\\ \cmidrule{2-3} \cmidrule{5-6} \cmidrule{8-9}
steps & mean & std &&  mean & std &&  mean & std \\ \midrule
GDP    \\
1 & 28.01 & 41.46 && 46.37 & 50.81 && 51.02 & 56.58 \\
 2 & 61.58 & 79.01 && 79.97 & 97.29 && 98.21 & 110.71 \\
 3 & 91.24 & 116.77 && 105.50 & 124.53 && 120.59 & 148.80 \\
 4 & 119.06 & 147.81 && 120.42 & 141.74 && 135.97 & 167.98 \\
 5 & 155.03 & 166.91 && 133.42 & 158.50 && 150.30 & 180.91 \\
ER  \\
1 & 4.99 & 7.37 && 4.67 & 4.69 && 5.48 & 5.37 \\
 2 & 6.16 & 5.95 && 6.00 & 5.69 && 7.50 & 6.97 \\
 3 & 6.96 & 5.95 && 6.65 & 4.61 && 7.19 & 5.75 \\
 4 & 6.77 & 6.77 &&6.52 & 5.49 && 6.39 & 6.52 \\
 5 & 7.48 & 7.92 && 6.91 & 6.05 && 7.36 & 6.39 \\ \bottomrule
\end{tabular}
\caption{\label{error_an} Russia. Mean and standard deviation of absolute error of out-of-sample forecasts for the proposed method, TVP-VAR \eqref{eq:gdp_exo-1} with exogenous variables and VAR.}
\end{table*}

\FloatBarrier

\section{Conclusions}

In the paper we propose a  TVP-VAR model with a time-invariant constraint on the long-run  multipliers of endogenous variables with respect to changes in exogenous variable. We provide  a Bayesian estimation method for TVP-VAR parameters and multipliers.
The proposed methodology can be used for  a wide range of practical applications as an alternative to VARX,  for example, in open economies modeling. Our approach is tailored to  economic systems whose cross-correlation relationships change due to changes in the monetary policy and exchange rate regimes under the hypothesis of long-run money neutrality. In the modern New Keynesian models the particular monetary policy rule  matters 
%
 for the shape of transition path from one long-run equilibrium to another. However, usually there is no influence of the monetary policy rule on the long-run equilibrium.
We apply the proposed methodology to model relationship  between the real GDP, the real exchange rate and real oil prices for the Norwegian and the Russian economies. Results show that incorporating the time invariance constraint for the long-run multipliers significantly improves forecasting performance of TVP-VAR model. Impulse responses are interpretable. The oil price increase leads to statistically significant real exchange rate appreciation and GDP increase in long run.   During periods of gradual reaction of the real exchange rate to the oil price shocks, real GDP reacted strongly to the shock. During periods of the sharp reaction of the real exchange rate, the real GDP demonstrates gradual increase. Therefore our results are in line with classical view that flexible exchange rates are shock absorbers for small open economies and the  floating exchange rate regime of monetary policy reduces volatility of the GDP growth.

\section*{References}
\FloatBarrier
\bibliographystyle{elsarticle-harv}
\bibliography{lit}

\end{document}